\documentclass[a4paper,12pt]{article}
\usepackage{amsfonts}
\usepackage{latexsym}
\usepackage{amssymb}
\usepackage{color}
\date{\empty}

\begin{document}
%\tableofcontents

\title{
Second Law and Liu Relations:\\ 
The No-reversible-direction Axiom - Revisited
}
\author{W. Muschik\footnote{Corresponding author:
muschik@physik.tu-berlin.de}
\\
Institut f\"ur Theoretische Physik\\
Technische Universit\"at Berlin\\
Hardenbergstr. 36\\D-10623 BERLIN,  Germany}
\maketitle

            \newcommand{\be}{\begin{equation}}
            \newcommand{\beg}[1]{\begin{equation}\label{#1}}
            \newcommand{\ee}{\end{equation}\normalsize}
            \newcommand{\bee}[1]{\begin{equation}\label{#1}}
            \newcommand{\bey}{\begin{eqnarray}}
            \newcommand{\byy}[1]{\begin{eqnarray}\label{#1}}
            \newcommand{\eey}{\end{eqnarray}\normalsize}
            \newcommand{\beo}{\begin{eqnarray}\normalsize}
         
            \newcommand{\R}[1]{(\ref{#1})}
            \newcommand{\C}[1]{\cite{#1}}

 \newcommand{\pddt}{\mbox{$\displaystyle
                                \frac{\partial}{\partial t}$}}
            \newcommand{\ddt}{\mbox{$\displaystyle
                                \frac{d}{d t}$}}
            \newcommand{\pddx}{\frac{\partial}{\partial\mvec{x}}}

            \newcommand{\mvec}[1]{\mbox{\boldmath{$#1$}}}
            \newcommand{\x}{(\!\mvec{x}, t)}
            \newcommand{\m}{\mvec{m}}
            \newcommand{\F}{{\cal F}}
            \newcommand{\n}{\mvec{n}}
            \newcommand{\argm}{(\m ,\mvec{x}, t)}
            \newcommand{\argn}{(\n ,\mvec{x}, t)}
            \newcommand{\T}[1]{\widetilde{#1}}
            \newcommand{\U}[1]{\underline{#1}}
            \newcommand{\V}[1]{\overline{#1}}
            \newcommand{\ub}[1]{\underbrace{#1}}
            \newcommand{\X}{\!\mvec{X} (\cdot)}
            \newcommand{\cd}{(\cdot)}
            \newcommand{\Q}{\mbox{\bf Q}}
            \newcommand{\p}{\partial_t}
            \newcommand{\z}{\!\mvec{z}}
            \newcommand{\bu}{\!\mvec{u}}
            \newcommand{\rr}{\!\mvec{r}}
            \newcommand{\w}{\!\mvec{w}}
            \newcommand{\g}{\!\mvec{g}}
            \newcommand{\D}{I\!\!D}
            \newcommand{\se}[1]{_{\mvec{;}#1}}
            \newcommand{\sek}[1]{_{\mvec{;}#1]}}            
            \newcommand{\seb}[1]{_{\mvec{;}#1)}}            
            \newcommand{\ko}[1]{_{\mvec{,}#1}}
            \newcommand{\ab}[1]{_{\mvec{|}#1}}
            \newcommand{\abb}[1]{_{\mvec{||}#1}}
            \newcommand{\td}{{^{\bullet}}}
            \newcommand{\eq}{{_{eq}}}
            \newcommand{\eqo}{{^{eq}}}
            \newcommand{\f}{\varphi}
            \newcommand{\rh}{\varrho}
            \newcommand{\dm}{\diamond\!}
            \newcommand{\seq}{\stackrel{_\bullet}{=}}
            \newcommand{\sta}[2]{\stackrel{_#1}{#2}}
            \newcommand{\om}{\Omega}
            \newcommand{\emp}{\emptyset}
            \newcommand{\bt}{\bowtie}
            \newcommand{\btu}{\boxdot}
            \newcommand{\tup}{_\triangle}
            \newcommand{\tdo}{_\triangledown}
            \newcommand{\Ka}{\frac{\nu^A}{\Theta^A}}
            \newcommand{\K}[1]{\frac{1}{\Theta^{#1}}}
            \newcommand{\ap}{\approx}
            \newcommand{\bg}{\sta{\Box}{=}}
            \newcommand{\si}{\simeq}
            \newcommand{\GG}{\mvec{]}}
            \newcommand{\LL}{\mvec{[}}
            \newcommand{\bx}{\boxtimes}
\newcommand{\Section}[1]{\section{\mbox{}\hspace{-.6cm}.\hspace{.4cm}#1}}
\newcommand{\Subsection}[1]{\subsection{\mbox{}\hspace{-.6cm}.\hspace{.4cm}
\em #1}}

\newcommand{\const}{\textit{const.}}
\newcommand{\vect}[1]{\underline{\ensuremath{#1}}}  %Vektoren
\newcommand{\abl}[2]{\ensuremath{\frac{\partial #1}{\partial #2}}}

\noindent
{\bf Abstract:} A thermodynamic process is governed by balance equations in field-formulated
thermodynamics. Especially the balance equation of entropy takes a prominent role: it introduces
the Second Law in the form of a dissipation inequality via the non-negative entropy production.
Balance equations and dissipation inequality are independent of the considered material which is
described by additional constitutive equations which need the introduction of a state space.
Inserting these constitutive equations into the balance equations results in the balance equations on
state space which include process directions. Why do not appear the latter in the Liu Relations
which describe material as the equations on state space do ? The answer is given by the axiom:
process directions which are in the kernel of the balance equations on state space do not enter the entropy production. A toy example concerning heat conduction in compressible fluids is added for
elucidation. 
\vspace{.6cm}

\section{Introduction}
\subsection{Basic concepts\label{BC}}

Non-equilibrium thermodynamics in field formulation\footnote{statistical and stochastic thermodynamics are out of scope here} launches out with {\em balance
equations}\footnote{the other description is: thermodynamics of discrete systems}\C{GM,PAP,IM}.
The minimal concept consists of the four balance equations for spinless materials
\byy{b1}
\mbox{mass:}&&\ d_t\varrho+\varrho\nabla\mvec{v}:{\bf 1}\ =\ 0,
\\ \label{b2}
\mbox{momentum:}&&\ \varrho d_t\mvec{v}+\nabla\cdot{\bf P}-\mvec{k}\ =\ \mvec{0},\ \
{\bf P}^\top ={\bf P},
\\ \label{b3}
\mbox{internal energy:}&&\ \varrho d_t\varepsilon+\nabla\cdot\mvec{q}+\nabla\mvec{v}:{\bf P}
-r\ =\ 0,
\\ \label{b4}
\mbox{entropy:}&&\ \varrho d_t s+\nabla\cdot\mvec{\Phi}-\gamma-\sigma\ =\ 0,\ \
\sigma\geq 0,
\eey
(mass density $\varrho$, material velocity $\mvec{v}$, symmetric pressure tensor ${\bf P}$,
external force density $\mvec{k}$, specific internal energy $\varepsilon$, heat flux density
$\mvec{q}$, energy supply $r$, specific entropy $s$, entropy flux density $\mvec{\Phi}$, entropy
supply $\gamma$, non-negative entropy production $\sigma$,
$d_t := \partial_t +\mvec{v}\cdot\nabla$)\footnote{{$^\top$means transposed, a:=b marks a
setting by definition, a is defined by b.}}.

The entropy balance equation \R{b4}$_1$ results by taking the 2nd law \R{b4}$_2$ into account
in a {\em dissipation inequality}
\bee{b5}
\varrho d_t s+\nabla\cdot\mvec{\Phi}-\gamma\ \geq\ 0,
\ee
and the question arises, how to achieve the compatibility of the balance equations \R{b1} to \R{b3}
with the dissipation inequality \R{b5}?

According to \R{b1} to \R{b3}, the wanted (basic) fields span the (large) {\em state space}
\C{MuPaEh,MuPaEh96}
\bee{b6}
{\cal Z} \ni\mvec{z} := (\varrho,\mvec{v},\varepsilon).
\ee
The other quantities in \R{b1} to \R{b3} and \R{b5} are constitutive fields which depend on the
basic fields
\bee{b7}
{\bf M} := ({\bf P},\mvec{q},s,\mvec{\Phi})(\mvec{z})
\ee
or are given by the system's environment
\bee{b8}
{\bf N} := (\mvec{k},r,\gamma).
\ee
Consequently, the differential operators $d_t$ and $\nabla$ act on a constitutive field $\boxplus$
by applying the chain rule
\bee{b9}
\nabla\cdot\boxplus = \frac{\partial\boxplus}{\partial\mvec{z}}:\nabla\mvec{z},\quad
d_t\boxplus=\frac{\partial\boxplus}{\partial\mvec{z}}\cdot d_t\mvec{z}.
\ee
Inserting \R{b9} into the balance equations \R{b1} to \R{b3} results in the so-called {\em balance
equations on the state space} \C{Mu23}.

A special non-systematical derived example of the dissipation inequality \R{b5} is the
{\em Clausius-Duhem inequality} \C{Tru,CL} which results from the setting
\bee{b10}
\mvec{\Phi}:=\frac{\mvec{q}}{\Theta},\qquad\gamma:=\frac{r}{\Theta}
\ee
(temperature\footnote{$\Theta$ may be the local equilibrium temperature by presupposing local
equilibrium or a non-equilibrium temperature such as the contact temperature
\C{CasVaz,MU14,MU77,MUBR}.} $\Theta$) which connects constitutive fields of the dissipation inequality
\R{b5} with those of the internal energy balance \R{b3} by physical intuition.

\subsection{{\bf Toy example:} Heat conducting fluid\label{TSV}}

The balance equations \R{b1} to \R{b4} do not contain the temperature, and consequently, the
temperature is not a state space variable \R{b6}. Introducing the specific free energy
\bee{b11}
\psi:=\varepsilon-\Theta s,
\ee
the internal energy balance \R{b3} becomes\footnote{$[\otimes]^s$ is the symmetric
part of $\otimes$} 
\bee{b12}
 \varrho d_t\psi+\varrho\Theta d_ts+\varrho s d_t\Theta+
\nabla\cdot\mvec{q}+[\nabla\mvec{v}]^s:{\bf P}-r\ =\ 0.
\ee
By the setting \R{b11},
the temperature $\Theta$ becomes a state variable, whereas the specific free energy $\psi$
and the specific entropy $s$ are constitutive equations on the state space replacing the internal
energy $\varepsilon$.

Taking the dissipation inequality \R{b5} into accout, \R{b12} results in an other shape of the
dissipation inequality which refers in contrast to \R{b5} not to the entropy density but to the
internal energy density
\bee{b13}
\varrho d_t\psi-\Theta\nabla\cdot\mvec{\Phi}+\Theta\gamma+\varrho s d_t\Theta+
\nabla\cdot\mvec{q}+[\nabla\mvec{v}]^s:{\bf P}-r \leq 0.
\ee
Multiplication with $1/\Theta$ generates a dissipation inequality with regard to the entropy density
\bee{b14}
\frac{\varrho}{\Theta}(d_t\psi+s d_t\Theta)
+\nabla\cdot(\frac{\mvec{q}}{\Theta}-\mvec{\Phi})
-\mvec{q}\cdot\nabla\frac{1}{\Theta}+(\gamma-\frac{r}{\Theta})
+\frac{1}{\Theta}[\nabla\mvec{v}]^s:{\bf P} \leq 0.
\ee
Introducing expressions
\bee{b14a}
\mvec{\Omega}\ :=\ \frac{\mvec{q}}{\Theta}-\mvec{\Phi},\qquad\kappa\ :=\ \gamma-\frac{r}{\Theta}
\ee
which are zero, if the Clausius-Duhem relations \R{b10} for the entropy flux density $\mvec{\Phi}$
and the entropy supply $\gamma$ are valid. The dissipation inequality \R{b14} becomes
\bee{b15}
-\frac{\varrho}{\Theta}(d_t\psi+s d_t\Theta)-\nabla\cdot\mvec{\Omega}
+\mvec{q}\cdot\nabla\frac{1}{\Theta}-\kappa
-\frac{1}{\Theta}[\nabla\mvec{v}]^s:{\bf P} \geq 0.
\ee
The state space and the constitutive fields which belong to \R{b15} and \R{b12} are
\bee{b16}
\mvec{z}=(\varrho,\Theta,\nabla\Theta,[\nabla\mvec{v]^s}), \quad
{\bf M}=(\psi,s,\mvec{q},{\bf P},r,\mvec{\Phi},\gamma).
\ee

Before this example is continued in sect.\ref{EC1}, some general remarks on the program and on
the balance equations on state space in matrix formulation are made.

\subsection{The program}

Now the question arises again, how to solve the balance equations on state space [\R{b1} to \R{b3}
with \R{b9} inserted]
by taking into consideration one of the dissipation inequalities \R{b5} or \R{b14} \C{MuTriPa}?
Using the cumbersome procedure of first solving the balance equations and then checking up whether
the dissipation inequality is satisfied at each position for all times, or better generating a
{\em class of materials} \C{MuPaEh} which always satisfies
the dissipation inequality by construction? Here, the second case is considered by using an easy, but
general toy-model which lightens the understanding.

Because the differential operators of the balance equations on state space are linear, the resulting
balance equations are linear in the so-called {\em higher derivatives} 
$\mvec{y}^\top = (d_t\mvec{z},\nabla\mvec{z})$ which are time and position
derivatives of the state space variables $\mvec{z}$
which are not included in the state space itself \C{MuPaEh}. Consequently, the
higher derivatives represent {\em process directions} in the state space. By using the evident, but
nevertheless axiomatic statement\vspace{.1cm}

{\em "process directions} 
{\em which are in the kernel (null-space) of the balance equations (on 

state space)}
{\em do not appear in the entropy production"} \footnote{more details in sect.\ref{EP}}, \vspace{.1cm}\\
the {\em no-reversible-directions axiom} can be established \C{Mu23,MuEh}:\vspace{.1cm}

{\em the local entropy production does not depend on the process directions.}\vspace{.1cm}\\
These statements allow to introduce a process independent {\em Lagrange multiplier}
which connects the balance equations directly to the dissipation inequality resulting in the
{\em Liu Relations} which do not include the process direction anymore \C{Li}.
The Liu Relations re\-present constraints on the constitutive equations with regard to the
dissipation inequality which can be exploited without taking process directions into account.
How this sketched program works in detail is performed below by considering a set of general
balance equations and the special Clausius-Duhem example presented in sect.\ref{BC}.

\section{Balance Equations on State Space}

Local {\em balance equations} can be written in different shapes
\bee{2}
\pddt (\rho \Psi )+\nabla \cdot (\rho \mvec{v}\Psi
+\mvec{\Phi})-\Sigma =0\ \longrightarrow\ 
\partial _{t}u_{A} + \partial _{j}\Phi^{j}_{A} = r_{A},\qquad
A = 1, 2, ....,M.
\ee
Here $\Psi$ or $u_A$ is the balanced basic field which may be of arbitrary rank,
 $\rho$ the mass density, $\mvec{\Phi}$ is the non-convective flux of $\Psi$ and $\Phi^j_A$
the total flux of $u_A$, $\mvec{v}$ the material velocity and $\Sigma$ or $r_A$ the sum of
production and supply of $\Psi$ or $u_A$, respectively. Among the $M$ equations is that of the
entropy $A:=S$
\bee{3}
u_S\ =\ \rho s,\quad \Phi^j_S\ =\ v^j\rho s +\Xi^j,\quad r_S\ =\
\gamma + \sigma\ \geq\ \gamma
\ee
(specific entropy $s$, non-convective entropy flux $\Xi^j$, entropy supply density $\gamma$,
entropy production density $\sigma \geq 0$). The inequality \R{3}$_4$ characterizes the Second
Law.

The fields in \R{2}$_2$ depend on the considered material which is characterized by a {\em state
space} ${\cal Z} \ni \mvec{z}$
{which is}
spanned by its components $\mvec{z}$.
These material dependent fields are the {\em constitutive equations}
\bee{4}
u_A\ =\ u_A(\mvec{z}),\quad \Phi^j_A\ =\ \Phi^j_A(\mvec{z}),\quad r_A\ =\ r_A(\mvec{z})
\ee
{which are inserted into the balance equations \R{2}$_2$ and the {\em dissipation
inequality} \R{3}$_4$}.
Applying the chain rule results in the so-called {\em {balance
{equations} on state space}} \R{5} and the {\em dissipation equation on state space} \R{6}
\C{MuPaEh,MuPaEh96}
\byy{5}
\frac{\partial u_A(\mvec{z})}{\partial\mvec{z}}\partial_t\mvec{z}+\frac{\partial \Phi^j_A(\mvec{z})}{\partial\mvec{z}}\partial_j\mvec{z} &=& r_A(\mvec{z}),\\ \label{6}
\frac{\partial u_S(\mvec{z})}{\partial\mvec{z}}\partial_t\mvec{z}+\frac{\partial \Phi^j_S(\mvec{z})}{\partial\mvec{z}}\partial_j\mvec{z} &\geq& \gamma(\mvec{z})
\eey
which are linear in the {\em higher derivatives}
\bee{7}
\mvec{y}^\top\ :=\ \Big(\partial_t\mvec{z},\partial_j\mvec{z}\Big).
\ee
Introducing the matrices
\byy{8}
{\bf A}(\mvec{z})\ :=\ \Big(\frac{\partial u_A(\mvec{z})}{\partial\mvec{z}},
\frac{\partial \Phi^j_A(\mvec{z})}{\partial\mvec{z}}\Big),&\quad&
\mvec{B}(\mvec{z})\ :=\ \Big(\frac{\partial u_S(\mvec{z})}{\partial\mvec{z}},
\frac{\partial \Phi^j_S(\mvec{z})}{\partial\mvec{z}}\Big),\\ \label{9}
\mvec{C}(\mvec{z})\ :=\ r_A(\mvec{z}),\hspace{2.5cm} 
&\quad& D(\mvec{z})\ :=\ \gamma(\mvec{z}),
\eey
the balances on state space \R{5} and \R{6} write
\bee{10}
{\bf A}(\mvec{z})\mvec{y}\ =\ \mvec{C}(\mvec{z}),\qquad
\mvec{B}(\mvec{z})\mvec{y}\ \geq\ D(\mvec{z}).
\ee
The wanted basic fields are $u_A\Big(\mvec{z}(t,\mvec{x})\Big)$, whereas
$\Phi_A^j\Big(\mvec{z}(t,\mvec{x})\Big)$ and $r_A\Big(\mvec{z}(t,\mvec{x})\Big)$
are given as functions of the basic fields or from the environment of the considered system.

\section{Example Continued: Higher derivatives\label{EC1}}

The balance equations \R{b1} and \R{b2} and the entropy balance \R{b4} are now transformed
into the matrix shape \R{10} which is linear in the following eight higher derivatives
\bee{c1}
\mvec{y}^\top\ =\ \Big( d_t\varrho,\nabla\varrho,d_t\mvec{v},d_t\Theta,d_t\nabla\Theta,
\nabla\nabla\Theta,d_t[\nabla\mvec{v}]^s,\nabla[\nabla\mvec{v}]^s\Big)
\ee
which follow from the linear balance equations and from the state space \R{b16}$_1$. Thus,
$\mvec{y}$ is a (8,1)-matrix and consequently, $\bf A$ is a (2,8)-matrix because two balances, \R{b1} and
\R{b2}, are taken into account, whereas the balance of the specific free energy \R{b12} is taken
into consideration by the dissipation inequality \R{b15} as demonstrated by replacing $d_ts$ in
\R{b12}. Because the dissipation inequality is scalar, $\mvec{B}$ is a (1,8)-matrix according to
\R{10}$_2$.

In detail, the matrix formulation of the balance equations \R{b1} and \R{b2} is
\bey\nonumber 
{\bf A}\mvec{y}:=\hspace{10cm} \\
=\left(
\begin{array}{cccccccc}
1 & 0 & 0 & 0 & 0 & 0 & 0 & 0 \\
0 & (\partial {\bf P}/\partial\varrho) & \varrho & 0 & 0 & (\partial {\bf P}/\partial\nabla\Theta) &
0 & (\partial {\bf P}/\partial[\nabla\mvec{v}]^s) \hspace{-.2cm} \\
\end{array}
\right)\mvec{y}\!\!\!&=&\!\!\!
\left(\!\!\!
\begin{array}{c}
-\varrho[\nabla\mvec{v}]^s:{\bf 1} \\
-(\partial {\bf P}/\partial\Theta)\cdot\nabla\Theta +\mvec{k} \hspace{-.3cm} \\
\end{array}\right)\nonumber\\ \label{c2}
&=:&\mvec{C}
\eey
and that of the dissipation inequality \R{b15}
\bey\nonumber
&&\small\mvec{B}\mvec{y}:=\\
&&\small-\!\!
\begin{array}{cccccccc}\Big(\!  %\hspace{-.2cm}
\frac{\varrho}{\Theta}(\partial\psi/\partial\varrho)\! &\! \partial\mvec{\Omega}/\partial\varrho
\!& 0\! & \frac{\varrho}{\Theta}[\partial\psi/\partial\Theta\!+\! s]\! &\!
\frac{\varrho}{\Theta}(\partial\psi/\partial\nabla\Theta)\! &\! \partial\mvec{\Omega}/\partial\nabla\Theta
\!&\! \frac{\varrho}{\Theta}(\partial\psi/\partial[\nabla\mvec{v}]^s)\! &\! \partial\mvec{\Omega}/\partial[\nabla\mvec{v}]^s
\!\Big)
\end{array}\!\!\!
\mvec{y}\hspace{.3cm}
\nonumber \\ 
&&\small\hspace{4cm}\geq\Big(\partial\mvec{\Omega}/\partial\Theta
-\mvec{q}\Big)\cdot\nabla\frac{1}{\Theta} + \kappa +\frac{1}{\Theta} [\nabla\mvec{v}]^s:{\bf P}=:D.
\label{c3}
\eey
$\mvec{C}$ is a (2,1)-matrix and $D$ a (1,1)-matrix.

Before the example can be continued in sect.\ref{EC2}, some more general considerations
are needed.

\section{Material Axioms}

The constitutive equations
{\R{4} generating}
the balance equations on state space \R{10}
cannot be arbitrary.
${\bf A}(\mvec{z})$, $\mvec{C}(\mvec{z})$, ${\mvec{B}}(\mvec{z})$ and $D(\mvec{z})$
have to satisfy material axioms which describe constraints which materials have to satisfy.
These axioms are:\\
1. The Second Law \R{10}$_2$ and its additionals,\\
2. Transformation properties by changing the observer,\\
3. Material symmetry,\\
4. State spaces $\cal Z$ that guarantee finite speed of wave propagation.\\ 
Here, we are interested in \#1. More hints can be found in the literature \C{PAP,MuPaEh,TrNo,TrTou}.

\section{Process Directions}

The state space variables $\mvec{z}(t,\mvec{x})$ depend on time and position according to \R{5}
and \R{6}. For each event $(t,\mvec{x})$, the higher derivatives \R{7} represent a
{\em process direction} in the state space $\cal Z$. Having solved the balance equations
\R{10}$_1$ (or \R{5}) for given initial conditions and geometrical constraints,
we know the higher derivatives $\mvec{y}^\top(t,\mvec{x})$, and we have to check,
whether or no these higher derivatives satisfy the dissipation inequality \R{10}$_2$.
Only if it is satisfied, the balance equations \R{10}$_1$ can be
attached to a thermodynamical process, depending on the chosen constitutive equations
${\bf A}(\mvec{z})$, ${\mvec C}(\mvec{z})$, $\mvec{B}(\mvec{z})$ and $D(\mvec{z})$.
This procedure is cumbersome, because if the dissipation inequality \R{10}$_2$ is not satisfied
by the chosen constitutive equations, we have to start the procedure again with renewed
constitutive equations until \R{10}$_2$ is satisfied. This is the so-called {\em global procedure}
which requires the solution of the balance equations.

Does a more specific procedure exist which can determine possible constitutive equations
in advance without solving \R{10}$_1$ {so that} \R{10}$_2$ is satisfied ? To
answer this question, the space of the higher derivatives (of all process directions) at an
arbitrary, but fixed event ${(t_0,\mvec{x}_0)}$ is introduced
\bee{11}
{\cal Y}(t_0,\mvec{x}_0)\ \ni\ \mvec{y}^\top (t_0,\mvec{x}_0).
\ee
That means, the global consideration of the balance equations \R{10}$_1$ is changed into a
{\em local procedure} at arbitrary, but fixed $(t_0,\mvec{x}_0)$. The balance equations \R{10}
are replaced by a local set of identical balance equations with different values of the higher
derivatives
\byy{h1}
&&\mvec{y}^{k\top}\in{\cal Y}(t_0,\mvec{x}_0):\\ \label{h2}
&&{\bf A}\Big(\mvec{z}(t_0,\mvec{x}_0)\Big)\mvec{y}^k = 
\mvec{C}\Big(\mvec{z}(t_0,\mvec{x}_0)\Big),\quad 
\mvec{B}\Big(\mvec{z}(t_0,\mvec{x}_0)\Big)\mvec{y}^k \geq 
D\Big(\mvec{z}(t_0,\mvec{x}_0)\Big),
\eey
searching for higher derivatives $\mvec{y}^k$ which satisfy \R{h2} at $(t_0,\mvec{x}_0)$ and then
along the complete process $(t,\mvec{y}(t))$. 

For the following, two statements (I and II) are considered which exclude each other \C{Mu23}:\\ \\
I. All local solutions\footnote{irreversible and reversible ones} of the local balance equations
{\R{h2}$_1$} satisfy the dissipation inequality {\R{h2}$_2$}  
\byy{12}
\bigwedge_k\Big\{\mvec{y}^k\in{\cal Y}_>(t_0,\mvec{x}_0)|{\bf A}\mvec{y}^k=\mvec{C}\Big\}
&\longrightarrow&
\Big\{\mvec{B}\mvec{y}^k> D\Big\},\\ \label{12a}
\bigwedge_m\Big\{\mvec{y}^m_{rev}\in{\cal Y}_=(t_0,\mvec{x}_0)|
{\bf A}\mvec{y}^m_{rev}=\mvec{C}\Big\}
&\longrightarrow&
\Big\{\mvec{B}\mvec{y}^m_{rev} = D\Big\}.
\eey
The process directions are divided into irreversible $\mvec{y}^k$ and reversible
$\mvec{y}^m_{eq}$ ones. The linear combined process direction,
{1}$>\alpha>$0,
\bee{12b}
{\bf A}\Big(\alpha\mvec{y}^k+(1-\alpha)\mvec{y}^m_{eq}\Big)\ =\ \mvec{C}\ 
\longrightarrow\ \mvec{B}\Big(\alpha\mvec{y}^k+(1-\alpha)\mvec{y}^m_{eq}\Big)\ >\ D
\ee
satisfies the balance equations and belongs to an irreversible process according to \R{12b}$_2$ and
\R{12}$_2$. No additional reversible process directions belonging to the subspace
${\cal Y}_=(t_0,\mvec{x}_0)\subset{\cal Y}(t_0,\mvec{x}_0)$ can be created by a linear combination of
$\mvec{y^k}$ and $\mvec{y}^m_{eq}$.\\ \\ 
The second statement is:\\
II. According to \R{12} and \R{12a}, there are local solutions of the balance equations
(on state space) $\mvec{y}^k$ and $\mvec{y}^m_{eq}$ which satisfy the dissipation inequality. Additional process directions $\mvec{y}^j_\Box$ are now presupposed which do not
satisfy \R{12} and \R{12a} in common, representing local solutions of the balance equations which
do not satisfy the dissipation inequality {\R{h2}$_2$} and which therefore do not represent a
thermodynamical process
\bee{13}
\bigwedge_j\Big\{\mvec{y}^j_\Box\in{\cal Y}_<(t_0,\mvec{x}_0)|
{\bf A}\mvec{y}^j_\Box=\mvec{C}\Big\}\ \longrightarrow\ 
\Big\{\mvec{B}\mvec{y}^j_\Box< D\Big\}.
\ee

By presupposing \R{13}, the space of the higher derivatives \R{11} is
\bee{16} 
{\cal Y}(t_0,\mvec{x}_0)\ =\ {\cal Y}_>(t_0,\mvec{x}_0) \cup {\cal Y}_=(t_0,\mvec{x}_0)
\cup {\cal Y}_<(t_0,\mvec{x}_0).\\
\ee
As proved in the next section \ref{RPD}, from statement \#II follows that additional reversible
process directions beyond those in ${\cal Y}_= (t_0,\mvec{x}_0)$ can be created from
${\cal Y}_>(t_0,\mvec{x}_0)\cup{\cal Y}_<(t_0,\mvec{x}_0)$.
This strange result paves the way to the axiom of no-reversible process directions.

\section{Reversible Process Directions in Non-equilibrium?\label{RPD}}

From \R{12}$_2$ and \R{13}$_2$ follows
\bee{17}
\mvec{B}\mvec{y}^k > D > \mvec{B}\mvec{y}^j_\Box\quad\longrightarrow\quad
\mvec{B}\mvec{y}^k -D\ >\ 0\ >\ \mvec{B}\mvec{y}^j_\Box-D.
\ee
The inequality \R{17}$_2$ is transformed to an equality by introducing two positve constants
\bee{18} 
\alpha >0,\ \beta >0:\qquad
\mvec{B}\mvec{y}^k -D+\alpha(\mvec{B}\mvec{y}^j_\Box-D)\ \doteq\
\mvec{B}\mvec{y}^j_\Box-D+\beta(\mvec{B}\mvec{y}^k-D).
\ee
This results in
\bee{19}
\mvec{B}\mvec{y}^k(1-\beta)+\mvec{B}\mvec{y}^j_\Box(\alpha-1)\ =\ D(-\beta+\alpha).
\ee
Setting
\bee{20}
-\beta+\alpha\ \doteq\ 1\ \longrightarrow\ \alpha-1 = \beta
\ee
creates a reversible process direction
\bee{21}
\mvec{B}\Big((1-\beta)\mvec{y}^k+\beta\mvec{y}^j_\Box\Big)\ =\ D
\ee
which is according to \R{12}$_1$ and \R{13}$_1$ also a local solution of the balance equations
\bee{22}
{\bf A}\Big((1-\beta)\mvec{y}^k+\beta\mvec{y}^j_\Box\Big)\ =\ 
(1-\beta)\mvec{C}+\beta\mvec{C}\ =\ \mvec{C},\quad 0<\beta<1,\quad (2>\alpha >1).
\ee
Consequently, if not all local solutions of the balance equations satisfy the dissipation inequality
(\#II), a reversible process direction \R{21} can be constructed by linear combination of
non-reversible ones (${\cal Y}_>(t_0,\mvec{x}_0)\cup{\cal Y}_<(t_0,\mvec{x}_0)$) which
are a local solutions of the balance equations \R{22}. This strange result is discussed in the next section \ref{CM2}.

\section{Coleman-Mizel's Shape of the Second Law\label{CM2}}

The dissipation inequality \R{10}$_2$ representing the Second Law can be differently
interpreted:\vspace{.2cm}

(\#I) If all process directions satisfy the dissipation
inequality, arbitrary constitutive equations are not possible, they are restricted by the
dissipation inequality.

(\#II) If the constitutive equations ${\bf A}(\mvec{z})$, ${\mvec C}(\mvec{z})$, $\mvec{B}(\mvec{z})$ and $D(\mvec{z})$ are given, the dissipation inequality
{\R{10}$_2$}
excludes those process directions which
do not satify the balance equations \R{10}$_1$.\vspace{.2cm}

The Second Law states nothing about these two each other excluding cases. Consequently, an
amendment to the Second Law is required for deciding which case, \#I or \#II, is valid
\C{MuEh,CimRog}. Here, this decision is given by an axiom which excludes the situation described
in section \ref{RPD}:
\bey\nonumber
\mbox{\bf A reversible process direction\hspace{1.6cm}}\\ \label{23} 
\mbox{\bf cannot be generated by non-reversible ones.}
\eey
Therefore the second inequality of \R{17}$_1$ must not be valid, that means
\bee{24}
{\cal Y}_<(t_0,\mvec{x}_0)\ =\ \emptyset,\quad\mbox{for all }(t_0,\mvec{x}_0)
\ee
is valid and consequently
{process directions of negative entropy production do not exist.
Consequently, all process directions are of non-negative entropy production.
Especially, there are by choice of ${\bf A},\mvec{C},\mvec{B},D$\footnote{representing the choice
of an admissible material} no solutions of the balance equations of negative entropy production.
Therefore}
\#I, \R{12} and \R{12a}, is true
\bey\nonumber
&&\mbox{\bf All local solutions of the balance equations}\\ \label{25} 
&&\hspace{.4cm}\mbox{\bf have to satify the dissipation inequality}.
\eey

After having excluded 
{by \R{24} all process directions of negative entropy production},
the dissipation inequality
represents a constraint with regard to the constitutive equations \R{8} and \R{9} and is not
excluding any process direction of $\Big({\cal Y}_>(t_0,\mvec{x}_0)\cup
{\cal Y}_=(t_0,\mvec{x}_0)\Big)$ which satisfy the Second Law: Consequently,
the constitutive equations, cannot be independent of each other, but must have the
property, that the entropy production density is not negative for all local -and therefore
also for all global- solutions of the balance equations. This is the {\em Coleman-Mizel
(CM)-formulation of the Second Law} \C{CoMi} which presupposes the validity of
\R{12} and \R{12a} ad-hoc. Taking \R{24} into account, the CM-formulation of the Second Law
follows and adopts its physical interpretation: 
\bey\nonumber
&&\mbox{\bf The constitutive equations enforce that there are no solutions}\\ \label{25a}
&&\hspace{.7cm}\mbox{\bf of the balance equations of negative entropy production.} 
\eey

\section{No-reversible Process Direction Axiom\label{EP}}

Taking a special material into consideration for which the process-direction-independent 
${\bf A}, \mvec{C}, \mvec{B}$ and
$D$ are given and for which \R{25} is valid. According to \R{12b}, a lot of processes of different
process directions are possible, resulting in the fact that
the matrix $\bf A$ of the constitutive equations has a kernel ${\cal K}$,
$\Big({\cal Y}_>(t_0,\mvec{x}_0)\cup{\cal Y}_=(t_0,\mvec{x}_0)\Big)
\supset{\cal K}\ni\mvec{y}_{ker}$
\bee{26}
{\bf A}\mvec{y}_{ker}\ =\ \mvec{0}\quad\longrightarrow\quad\mvec{y}\ =\ \mvec{y}_0
+\mvec{y}_{ker}\quad\longrightarrow\quad{\bf A}\mvec{y}\ =\ 
{\bf A}\mvec{y}_0\ =\ \mvec{C}.
\ee
Here $\mvec{y}$ and $\mvec{y}_0$ are local solutions of the balance equations according to
\R{10}$_1$, and with them also
\bee{27}
{\bf A}\Big(\alpha\mvec{y}+(1-\alpha)\mvec{y}_0\Big)\ =\ \mvec{C}
\ee
satisfies the balance equation.

Introducing the entropy production $\sigma\geq 0$, the dissipation inequality becomes according
to \R{12} 
\bee{28}
\mvec{B}\mvec{y}\ =\ \mvec{B}(\mvec{y}_0+\mvec{y}_{ker})\ =\ 
D+\sigma(\mvec{y})\ =\ D+\sigma(\mvec{y}_0+\mvec{y}_{ker})
\geq\ D.
\ee
According to \R{26}$_1$, the kernel of ${\bf A}$ is not present in the balance equations and
therefore it should also not be present in the entropy production
\bee{29}
\framebox[6cm]{$
\sigma(\mvec{y})\ =\ \sigma(\mvec{y}_0+\mvec{y}_{ker})\ 
\sta{{\bf ax}}{=}\ \sigma(\mvec{y}_0).$}
\ee
Although being clear from a view of physics, \R{29}$_2$  is an axiom which is more stringent than
the verbal formulations \R{23} and \R{25} as demonstrated below at \R{34}.
Consequently, from \R{28}$_2$ and \R{29}$_{1,2}$ follows 
\bee{30}
\mvec{B}\mvec{y}-\mvec{B}\mvec{y}_0\ =\ D+
\sigma(\mvec{y})-(D+\sigma(\mvec{y}_0))\ =\ 0\ =\ 
\mvec{B}(\mvec{y}-\mvec{y}_0)\ =\ \mvec{B}\mvec{y}_{ker},
\ee
that means, $\mvec{B}$ is perpendicular to the kernel of ${\bf A}$.

Consider two arbitrary local solutions of the balance equations \R{26}$_3$, 
$\mvec{y}^1$ and $\mvec{y}^2$, and additionally \R{26}$_{1,2}$
\bee{31}
\mvec{C}-\mvec{C}\ =\ 
\mvec{0}\ =\ {\bf A}(\mvec{y}^1-\mvec{y}^2)\ =\  {\bf A}(\mvec{y}^1_0-\mvec{y}^2_0)\ 
\longrightarrow\ \mvec{y}^1_0-\mvec{y}^2_0\ =:\ \mvec{y}^{12}_{ker}\ \in\ \cal{K},
\ee
resulting according to \R{26}$_2$, \R{31}$_3$ and \R{30}$_4$ in
\byy{32}
\mvec{y}^1-\mvec{y}^2 &=& \mvec{y}^{12}_{ker}+\mvec{y}^1_{ker}-\mvec{y}^2_{ker},
\\ \label{33}
\mvec{B}(\mvec{y}^1-\mvec{y}^2) &=& 
\mvec{B}(\mvec{y}^{12}_{ker}+\mvec{y}^1_{ker}-\mvec{y}^2_{ker})\ =\ 0\ =\ 
\sigma(\mvec{y}^1)-\sigma(\mvec{y}^2),
\eey
according to \R{28}$_2$. Consequently, the entropy production does not locally depend on the
process direction according to \R{33}$_3$. If one process direction at $(t_0,\mvec{x}_0)$ is
reversible,\\ $\mvec{y}_{rev}\in{\cal Y}_=(t_0,\mvec{y}_0)$, all other process directions at
$(t_0,\mvec{x}_0)$ are also reversible, and an equilibrium state is present. This results in the
verbal formulation of the {\em Axiom of No-reversible Process Directions} \C{WM} which
represents an amendment of the Second Law:
\bey\nonumber
&&\hspace{3.4cm}\mbox{\bf Except in equilibria,}\\ \label{34}
&&\mbox{\bf reversible process directions in state space do not exist}
\eey

\section{Liu Relations}

Introducing a suitable vector $\mvec{\lambda}(\mvec{y})$ which for the present depends on the
process direction $\mvec{y}$, and by use of \R{28}$_2$ and \R{26}$_3$, the following
equality is valid
\bee{35}
(\mvec{B}-\mvec{\lambda}{\bf A})\mvec{y}\ =\ D+\sigma(\mvec{y})
-\mvec{\lambda}(\mvec{y})\mvec{C}.
\ee
Now $\mvec{\lambda}$ can be chosen specially, depending on the entropy production
\bee{g1}
\mvec{\lambda}(\mvec{y})\mvec{C}\ \doteq\ D+\sigma(\mvec{y}),
\ee
taking into account that $\mvec{C}$ and $D$ do not depend on the higher derivatives $\mvec{y}$.
Because the entropy production does not depend on the process direction according to \R{33}$_3$,
$\mvec{\lambda}$ is also independent of the process direction.

The choice of a process-direction-independent $\mvec{\lambda}$ in the scalar product of the
LHS of \R{g1} is not unequivocal, but always possible, and by taking \R{g1} into account 
\R{35} results in
\bee{g1a}
(\mvec{B}-\mvec{\lambda}{\bf A})\mvec{y}\ =\ D+\sigma-\mvec{\lambda}\mvec{C}\ =\ 0
\ee
which is valid for all higher derivatives resulting in
\bee{g2}
\framebox[2cm]{$
\mvec{B}\ =\ \mvec{\lambda}{\bf A}$}\quad\mbox{and}\quad
\framebox[2.2cm]{$
\mvec{\lambda}\mvec{C}\ \geq\ D$}. 
\ee
Thus, the process-direction-independent {\em Liu Relations} \R{g2} \C{Li,TriPapCimMu} are
generated by the setting \R{g1} and taking into account that the entropy production is independent
of the process direction.

Because ${\bf A}$ has no right-hand inverse, $\mvec{\lambda}$ is not determined by the given
$\mvec{B}$ and ${\bf A}$, but the constitutive quantities $\mvec{B}$ and ${\bf A}$ are not
independent of each other: there exists a $\mvec{\lambda}$, so that the Liu Relations
\R{g2} are satisfied. Because there are less balance equations than higher
derivatives\footnote{see: eqs.\R{b1} to \R{b3} (r=3 balances) in comparison with eq.\R{c1}
(c=8 higher derivatives)}, the types (number of rows $r$ and of columns $c$) of the following
matrices are
\bee{g1x}
{\bf A}: (r,c),\ r < c,\quad \mvec{y}: (c,1),\quad
D: (1,1),\quad\mvec{B}: (1,c),\quad\mvec{\lambda}: (1,r),\quad\mvec{C}: (r,1). 
\ee
A tutorial example concerning the use of the Liu Relations can be found in
\C{MuPaEh}.

The Liu Relations represent constraints on the constitutive properties. They
bring together constitutive quantities of the balance equation with those of the dissipation
inequality. These restrictions have to be inserted into the balance equations and the dissipation
inequality \R{10}, resulting by use of \R{g1a} in a differential equation which takes the
dissipation inequality into account
\bee{37}
\mvec{B}\mvec{y}\ =\ \mvec{\lambda}\mvec{C}.
\ee

An other kind of exploitation the dissipation inequality regards the $\mvec{\lambda}$ as a
Lagrange multiplier introducing the balance equations into the dissipation inequality \C{Li}.
The dissipation inequality \R{35}$_2$ can be replaced by
\bee{38}
\mvec{B}\mvec{\omega}-\mvec{\lambda}({\bf A}\mvec{\omega}-\mvec{C})\ =\ 
D+\sigma\ \geq\ D.
\ee
Here, $\mvec{\omega}$ is an arbitrary process direction which not necessarily has to satisfy the
balance equations and/or the dissipation inequality \R{12} and/or \R{12a}. The usual dictum is:
add the balance equations as constraints with a Lagrange factor to the dissipation inequality
resulting in arbitrary process directions.

\section{Example Continued:\\ Liu Relations + Entropy Production\label{EC2}}

The Liu Relations \R{g2} belonging to the example \R{c2} and \R{c3} are as follows
\byy{d1}
\mvec{\lambda}{\bf A}=\Big(
\begin{array}{cc}\lambda^1 & \lambda^2 \end{array}\Big)
\left(
\begin{array}{cccccccc}
1 & 0 & 0 & 0 & 0 & 0 & 0 & 0 \\
0 & (\partial {\bf P}/\partial\varrho) & \varrho & 0 & 0 & (\partial {\bf P}/\partial\nabla\Theta) &
0 & (\partial {\bf P}/\partial[\nabla \mvec{v}]^s) \hspace{-.2cm} \\
\end{array}
\right)=\nonumber\\
=\Big(
\begin{array}{cccccccc}
\lambda^1 & \lambda^2(\partial {\bf P}/\partial\varrho) & \lambda^2\varrho 
& 0 & 0 & 
\lambda^2(\partial {\bf P}/\partial\nabla\Theta) & 0 & \lambda^2(\partial {\bf P}/\partial[\nabla\mvec{v}]^s)
\end{array}
\Big)=\mvec{B}=\nonumber\\
\small-\!\!
\begin{array}{cccccccc}\Big(\! 
\frac{\varrho}{\Theta}(\partial\psi/\partial\varrho)\! &\! \partial\mvec{\Omega}/\partial\varrho
\!& 0\! & \frac{\varrho}{\Theta}[\partial\psi/\partial\Theta\!+\! s]\! &\!
\frac{\varrho}{\Theta}(\partial\psi/\partial\nabla\Theta)\! &\! \partial\mvec{\Omega}/\partial\nabla\Theta
\!&\! \frac{\varrho}{\Theta}(\partial\psi/\partial[\nabla\mvec{v}]^s)\! &\! \partial\mvec{\Omega}/\partial[\nabla\mvec{v}]^s
\!\Big),
\end{array}\label{d1}
\eey
and
\bey\nonumber
\mvec{\lambda}\mvec{C}= 
\Big(
\begin{array}{cc}\lambda^1 & \lambda^2 \end{array}\Big)
\left(
\begin{array}{c}
-\varrho[\nabla\mvec{v}]^s:{\bf 1} \\  \hspace{-.3cm}
-(\partial {\bf P}/\partial\Theta)\cdot\nabla\Theta +\mvec{k} \hspace{-.3cm} \\
\end{array}
\right)= -\lambda^1 \varrho[\nabla\mvec{v}]^s:{\bf 1}
-\lambda^2\Big((\partial{\bf P}/\partial\Theta)\cdot\nabla\Theta-\mvec{k}\Big)\geq \\
%=\mvec{\Phi}\cdot\nabla\Theta + [\nabla\mvec{v}]^s:{\bf P}\\
\geq D=\Big(\partial\mvec{\Omega}/\partial\Theta
-\mvec{q}\Big)\cdot\nabla\frac{1}{\Theta} + \kappa
+\frac{1}{\Theta} [\nabla\mvec{v}]^s:{\bf P}.\hspace{1.2cm}
\label{d4}
\eey

From the matrix elements (1,3) and (1,1) of $\mvec{\lambda}{\bf A}=\mvec{B}$ follows
\bee{d2}
\lambda^2\ =\ 0,\qquad \lambda^1\ =\ -\frac{\varrho}{\Theta}(\partial\psi/\partial\varrho).
\ee
The matrix elements (1,4), (1,5) and (1,7) of $\mvec{\lambda}{\bf A}=\mvec{B}$ result in
\bee{d3}
s\ =\ -(\partial\psi/\partial\Theta),\qquad 0\ =\ (\partial\psi/\partial\nabla\Theta),
\qquad 0\ =\ (\partial\psi/\partial[\nabla\mvec{v}]^s),
\ee
and by taking \R{d2}$_1$ into account, (1,2), (1,6) and (1,8) are
\bee{d3a}
0\ =\ \partial\mvec{\Omega}/\partial\varrho,\qquad 
0\ =\ \partial\mvec{\Omega}/\partial\nabla\Theta,\qquad
0\ =\ \partial\mvec{\Omega}/\partial[\nabla\mvec{v}]^s.
\ee
The entropy production follows from the dissipation inequality \R{d4} by taking \R{d2} into account
\bee{d3b}
\frac{\varrho^2}{\Theta}(\partial\psi/\partial\varrho)[\nabla\mvec{v}]^s:{\bf 1}
-\Big(\partial\mvec{\Omega}/\partial\Theta
-\mvec{q}\Big)\cdot\nabla\frac{1}{\Theta} - \kappa
-\frac{1}{\Theta} [\nabla\mvec{v}]^s:{\bf P}\ \geq\ 0.
\ee

Especially for {\em Clausius-Duhem materials} which satisfy
\bee{d3c}
\mvec{\Omega}\ =\ 0,\qquad \kappa\ =\ 0
\ee
according to \R{b14a}, the entropy production \R{d3b} becomes
\bee{d5}
\frac{1}{\Theta}[\nabla\mvec{v}]^s:\Big(\varrho^2\frac{\partial\psi}{\partial\varrho}{\bf 1}
-{\bf P}\Big)+\mvec{q}\cdot\nabla\frac{1}{\Theta}\ \geq\ 0.
\ee

The calculation above takes the balance equation of the internal energy \R{b12} only indirectly
into account by inserting the dissipation inequality \R{b5} into \R{b12}, thus generating a
special shape of the dissipation inequality \R{b15} which is used together with the balances of
mass \R{b1} and momentum \R{b2}. Now the question arises, how the only implicit use
of the energy balance equation \R{b12} restricts the result compared to its full consideration ?
This question is the item of the next section.

\section{Example Continued:\\ Liu Relations + Class of Materials }

Now the Liu Relations are generated by use of all balance equations \R{b1} to \R{b3} and the
dissipation inequality \R{b5}. 
The balance equation of the internal energy is now completely taken into account, that means,
the (2,8)-matrix ${\bf A}$ in \R{c2} has to be supplemented by \R{b12} in matrix formulation
resulting in a third row of \R{c2}, whereas the balance equations of mass and momentum are
taken into account by the first two rows. Thus, the balance equations write
${\bf A}^\ast\mvec{y}=\mvec{C}^\ast$, (3,8)(8,1)=(3,1).
The state space \R{b16}$_1$ is chosen as above in sect.\ref{TSV}
resulting in the set of higher derivatives \R{c1}. 

The matrix formulation of \R{b3} according to the chosen state space \R{b16}$_1$ is
\bey\nonumber
\begin{array}{cccccccc}
\Big(\varrho\partial\varepsilon/{\partial\varrho} &
\partial\mvec{q}/{\partial\varrho} & 0 & \varrho\partial\varepsilon/{\partial\Theta}
& \varrho\partial\varepsilon/{\partial\nabla\Theta}
 & \partial\mvec{q}/{\partial\nabla\Theta} & 
\varrho\partial\varepsilon/{\partial[\nabla\mvec{v}]^s}
& \partial\mvec{q}/{\partial[\nabla\mvec{v}]^s}\Big)%\mvec{y}
\end{array}\!\!\mvec{y}\ =\\ \label{d6}
=\ -\partial\mvec{q}/\partial\Theta\cdot\nabla\Theta - \nabla[\mvec{v}]^s:{\bf P}+r.\hspace{.7cm}
\eey
The first two rows of ${\bf A}^\ast$ and
$\mvec{C}^\ast$ can be found in \R{c2} and the third row in \R{d6}.
The dissipation inequality is now \R{b5} instead of \R{b15}. Its matrix formulation according to the
chosen state space \R{b16}$_1$ is
\bey\nonumber
\begin{array}{cccccccc}
\Big(\varrho\partial s/{\partial\varrho} &
\partial\mvec{\Phi}/{\partial\varrho} & 0 & \varrho\partial s/{\partial\Theta}
& \varrho\partial s/{\partial\nabla\Theta}
 & \partial\mvec{\Phi}/{\partial\nabla\Theta} & 
\varrho\partial s/{\partial[\nabla\mvec{v}]^s}
& \partial\mvec{\Phi}/{\partial[\nabla\mvec{v}]^s}\Big)%\mvec{y}
\end{array}\!\!\mvec{y}\ =\\ \label{d7}
=:\ \mvec{B}^\ast \mvec{y}\
\geq\ \gamma-\partial\mvec{\Phi}/\partial\Theta\cdot\nabla\Theta\ =:\ D^\ast.\hspace{.7cm}
\eey

The corresponding Liu Relations \R{g2} are
\bee{d8}
\mvec{\lambda}^\ast {\bf A}^\ast\ =\ \mvec{B}^\ast, \qquad
\mvec{\lambda}^\ast\mvec{C}^\ast\ \geq\ D^\ast
\ee
written in components as done in \R{d1} and \R{d4}, \R{d8}$_1$ yields
\byy{d9}
\lambda^{1\ast}+\lambda^{3\ast}\varrho\frac{\partial\varepsilon}{\partial\varrho}
&=& \varrho\frac{\partial s}{\partial\varrho},\\
\label{d10}
\lambda^{2\ast}\frac{\partial{\bf P}}{\partial\varrho}
+\lambda^{3\ast}\frac{\partial\mvec{q}}{\partial\varrho} &=&
\frac{\partial\mvec{\Phi}}{\partial\varrho},\\
\label{d11}
\lambda^{2\ast}\varrho &=& 0,\\
\label{d12}
\lambda^{3\ast}\varrho\frac{\partial\varepsilon}{\partial\Theta}&=&
\varrho\frac{\partial s}{\partial\Theta},\\
\label{d13}
\lambda^{3\ast}\varrho\frac{\partial\varepsilon}{\partial\nabla\Theta}&=&
\varrho\frac{\partial s}{\partial\nabla\Theta},\\
\label{d14}
\lambda^{2\ast}\frac{\partial{\bf P}}{\partial\nabla\Theta}
+\lambda^{3\ast}\frac{\partial\mvec{q}}{\partial\nabla\Theta}
&=&
\frac{\partial\mvec{\Phi}}{\partial\nabla\Theta},\\
\label{d15}
\lambda^{3\ast}\varrho\frac{\partial\varepsilon}{\partial[\nabla\mvec{v}]^s}&=&
\varrho\frac{\partial s}{\partial[\nabla\mvec{v}]^s},\\
\label{d16}
\lambda^{2\ast}\frac{\partial{\bf P}}{\partial[\nabla\mvec{v}]^s}
+\lambda^{3\ast}\frac{\partial\mvec{q}}{\partial[\nabla\mvec{v}]^s}
&=&
\frac{\partial\mvec{\Phi}}{\partial[\nabla\mvec{v}]^s},
\eey
and \R{d8}$_2$ becomes by taking \R{c2}, \R{d6} and \R{d7} into account
\bey\nonumber
-\lambda^{1\ast}\varrho[\nabla\mvec{v}]^s:{\bf 1}
-\lambda^{2\ast}\Big((\partial {\bf P}/\partial\Theta)\cdot\nabla\Theta -\mvec{k}\Big)
+\lambda^{3\ast}\Big(-\partial\mvec{q}/\partial\Theta\cdot\nabla\Theta
- [\nabla\mvec{v}]^s:{\bf P}+r\Big)\geq\\ \label{d17}
\geq   \gamma-\partial\mvec{\Phi}/\partial\Theta\cdot\nabla\Theta,\hspace{.6cm}
\eey
resulting in the entropy production by taking \R{d11} into account
\bey\nonumber
-\lambda^{1\ast}\varrho[\nabla\mvec{v}]^s:{\bf 1}
-\partial(\lambda^{3\ast}\mvec{q}-\mvec{\Phi})/\partial\Theta\cdot\nabla\Theta
+\mvec{q}(\partial\lambda^{3\ast}/\partial\Theta)\cdot\nabla\Theta-\hspace{2cm}
\\ \label{d18}
- \lambda^{3\ast}([\nabla\mvec{v}]^s:{\bf P})+\lambda^{3\ast}r-\gamma\geq 0
\eey

The Liu procedure does not determine $\lambda^{1*}$ and $\lambda^{3*}$. But physical
understanding yields according to \R{b14a}$_2$ that $\lambda^{3*}$ is a reciprocal temperature
and $\lambda^{1*}$ is then determined by \R{d9}. Consequently, the class of materials which
fit the balance equations \R{b1} to \R{b3} and the dissipation inequality \R{b4} is charactrized
by \R{d9} to \R{d18}. A special sub-family is the {\em Clausius-Duhem family} which is defined
by $\mvec{\Omega}=\mvec{0}$ and $\kappa=0$ in \R{b14a}. Consequently, the entropy
production belonging to this sub-family is according to \R{d18}, \R{b16}$_1$ and \R{b11}
\byy{d19}
-\Big(\varrho\frac{\partial s}{\partial\varrho}-\frac{1}{\Theta}\varrho
\frac{\partial\varepsilon}{\partial\varrho}\Big)\varrho[\nabla\mvec{v}]^s:{\bf 1}
-\mvec{q}\frac{1}{\Theta^2}\cdot\nabla\Theta
-\frac{1}{\Theta}[\nabla\mvec{v}]^s:{\bf P}&\geq& 0,
\\ \label{d20}
\Big(\frac{\partial\varepsilon}{\partial\varrho}-\Theta\frac{\partial s}{\partial\varrho}\Big)
\varrho^2\frac{1}{\Theta}[\nabla\mvec{v}]^s:{\bf 1}
+\mvec{q}\cdot\nabla\frac{1}{\Theta}
-\frac{1}{\Theta}[\nabla\mvec{v}]^s:{\bf P}&\geq& 0,
\\ \label{d20a}
\frac{1}{\Theta}[\nabla\mvec{v}]^s:\Big(\frac{\partial\psi}{\partial\varrho}\varrho^2{\bf 1}
-{\bf P}\Big)+\mvec{q}\cdot\nabla\frac{1}{\Theta}&\geq& 0.
\eey
A comparison with \R{d5} demonstrates that the abridged procedure in sect.\ref{EC2} yields the
same result for the entropy of Clausius-Duhem materials, a fact which was not expected.
Introducing the pressure $p\doteq(1/3)\mbox{Tr}{\bf P}$, the entropy production follows
from \R{d20}
\bee{d21}
{\bf P}\ =\ p{\bf 1}+\mvec{\Pi}\ \longrightarrow\ \sigma\ =\  \frac{1}{\Theta}(\nabla\cdot\mvec{v})
\Big(\frac{\partial\psi}{\partial\varrho}\varrho^2-p\Big)+
\mvec{q}\cdot\nabla\frac{1}{\Theta}
-\frac{1}{\Theta}[\nabla\mvec{v}]^s:{\bf \Pi}
\ \geq\ 0.
\ee
Consequently, the entropy production of incompressible Clausius-Duhem materials without internal
friction is those of pure heat conduction
\bee{d22}
\Big(\nabla\cdot\mvec{v}=0\Big)\ \wedge\ \Big({\bf \Pi}={\bf 0}\Big)\ \longrightarrow\
\sigma\ =\ \mvec{q}\cdot\nabla\frac{1}{\Theta}\ \geq\ 0,
\ee
a hint at the efficacy of the Liu Relations \R{d8}.

\section{Summary}

For describing materials, a state space is necessary on which the linear differential ope\-rators of the
balance equations\footnote{$d_t$ and $\nabla$} act by applying the chain rule. Thus, time
and position derivatives of the state space variables\footnote{the so-called higher derivatives}
appear in the balance equations which for discrimination are called {\em balances on state
space} and which are linear in the {\em higher derivatives} which can be interpreted as {\em process
directions} in state space. Because there are less ba\-lance equations than higher derivatives, these
ones are not determined by the balances on state space, rather higher derivatives exist forming
a kernel of the balances on state space which are linear in the higher derivatives. Due to this  
linearity, higher derivatives can be combined linearily representing different process directions.
According to the dissipation inequality, there are three kinds of process directions: irreversible,
reversible and not-satisfying\footnote{the dissipation inequality} ones.

The axiomatical statement "A reversible process direction cannot be generated by non-reversible
ones" results in the Coleman-Mizel formulation of the Second Law
"All local solutions of the balance equations (on state space) have to satisfy the dissipation
inequality". A second axiomatical statement " Process directions which are in the kernel of the
balance equations (on state space) do not enter the entropy production" results in the {\em Axiom
of No-reversible Process Directions} "Except in equilibria, reversible process directions in state
space do not exist", that is, reversible process directions appear only in equilibrium and the entropy
production in non-equilibrium does not depend on the process direction, it is a state function.
Consequently, reversible process directions do not exist in non-equilibrium.

If the entropy production were dependent on the process direction, the {\em Liu Relations}
could not be derived by the here presented procedure. These results are based on the obvious, but
axiomatic fact that process directions which do not appear in the balance equations also do not
enter the entropy production. Using this axiom, the Liu Relations can be derived from the balance
equations and the dissipation inequality. These relations are independent of the process directions
and connect the balance equations directly with the dissipation inequality, thus generating
constraints for the choice of the material. The advantage of the Liu procedure with respect to the
Coleman-Noll technique \C{CoNo,TriPapCimMu} is that the Liu Relations do not contain higher
derivatives which by using the Coleman-Noll technique have to be removed by setting the brackets 
to zero which are factors of the higher derivatives.\\

{\bf Ethical Approval:} not applicable

{\bf Funding:} not applicable

{\bf Availability of data and materials:} not applicable

\end{document}